\documentstyle[sprocl]{article}

\def\be{\begin{equation}}
\def\ee{\end{equation}}
\def\bea{\begin{eqnarray}}
\def\eea{\end{eqnarray}}
\begin{document}
\begin{flushright}
LAEFF 95/13 \\
FTUAM 95/..
\end{flushright}

\title{RENORMALIZATION GROUP CALCULATION OF THE GALAXY-GALAXY CORRELATION
FUNCTION}

\author{J. P\'EREZ-MERCADER$^*$, T. GOLDMAN$^{\dag}$,
 D. HOCHBERG $^{\ddag,*}$\footnote{Based on a talk presented by
D. Hochberg at the International Workshop on Elementary Particle Physics,
Valencia, Spain,
June 5 to 9, 1995.}
\\ AND R. LAFLAMME$^{\dag}$ }

\address{$^*$Laboratorio de Astrof\'isica Espacial y F\'isica Fundamental\\
 Apartado 50727, 28080 Madrid, Spain}

\address{$^{\dag}$Theoretical Division, Los Alamos National Laboratory\\
 Los Alamos, New Mexico 87545, USA}

\address{$^{\ddag}$Departamento de F\'isica Te\'orica, Universidad Aut\'onoma
de Madrid\\
 Cantoblanco, 28049  Madrid, Spain}


\maketitle\abstracts{The observable Universe is described by a
collection of equal mass galaxies
linked into a common unit by their mutual gravitational interaction. The
partition
function of this system is cast in terms of Ising model spin variables and maps
exactly onto a three-dimensional stochastic
scalar classical field theory. The full machinery
of the renormalization group and critical phenomena is brought to bear on this
field theory
allowing one to calculate the galaxy-to-galaxy correlation function, whose
critical
exponent is predicted to be between 1.530 to 1.862, compared to the
phenomenological value of
1.6 to 1.8}

It is an empirical fact that the Universe today is dominated by matter which is
in
gravitational interaction and is organized into spatial configurations
consisting of galaxies, groups of galaxies, superclusters, and
even larger structures \cite{Lick,APM}. It is also
true that one of the outstanding problems in modern
cosmology is to provide a quantitative description of this
structure from first principles \cite{Peebles}.
One of the standard ways in which structure, that is,
the departure from randomness and homogeneity,
is described, is by means of correlation functions.
These objects are a basic tool for unravelling structure in physical systems
ranging from gases and liquids to solids. Experimental input is typically
deduced from analysis of the scattering of neutrons or x-rays. By contrast, in
cosmology, this input is provided by direct observation.
{}From observation, the galaxy-to-galaxy correlation function
$(\xi_{gal}(r))$ is known to scale as $\xi(r) \sim r^{-\gamma}$, where
$\gamma \sim O(1.6\,
 {\rm to}\, 1.8)$, instead of $\gamma = 1$, which is what one would
naively expect \cite{footnote1}  for
a homogeneous distribution of matter in a three dimensional space.

Given these observational facts, we shall demonstrate
how to {\em calculate} the
exponent $\gamma$ from first principles \cite{Gang4}.
We take our cue by analogy with the well
established theory of liquids and interacting gases \cite{Goods},
where the relative arrangements of the atoms and molecules
(galaxies, in our case!) are accurately described in terms of the principles
and methods
of classical
statistical mechanics. We will apply the same techniques to a system made up by
many
galaxies in gravitational interaction, subject to fluctuations emerging from
the intrinsic
properties of the gravitational interaction \cite{Gang4}.
We will obtain the canonical partition
function for this system and compute the two-point function and its corrections
due to
these fluctuations.

Begin by considering the continuous mass density $\rho ({\bf r})$ describing
the
spatial distribution of galaxies. The departure from perfect homogeneity is
$\delta \rho ({\bf r}) = \rho ({\bf r}) - {\bar \rho}$, where $\bar \rho$
denotes the
average density. The galaxy-galaxy correlation function is defined
\cite{Peebles} as
$\xi_{Gal}({\bf r}_i - {\bf r}_j) = <\delta \rho({\bf r}_i)
\delta \rho({\bf r}_j)>/{\bar \rho}^2$,
the angular brackets denoting a
suitable averaging. This is the joint probability distribution function
for finding a galaxy at position ${\bf r}_i$ given that there is one at ${\bf
r}_j$.
In the
static, nonrelativistic and weak gravitational field limit,
(we discuss the effects of expansion later) the gravitational interaction
energy for this system is
\begin{equation}
H_{int} = -\frac{G}{2} \int \int d^3{\bf r}_i \,d^3{\bf r}_j\, \rho ({\bf r}_i)
 \frac{1}{|{\bf r}_i - {\bf r}_j|} \,\rho ({\bf r}_j),
\end{equation}
where the integral excludes the point ${\bf r}_i = {\bf r}_j$, and
the factor of $1/2$ appears because each galaxy is counted twice in the
integration. It is
reasonable to consider the gas of $N$ galaxies as made
up of discrete spatially localized points of equal mass $m_o$ at position
${\bf r}_j$, so that
$\rho ({\bf r}) = \sum_{j = 1}^{N} m_j \delta ({\bf r} - {\bf r}_j)$. The
interaction
energy then becomes
\begin{equation}
H_{int} = \frac{-1}{2}\sum_{i,j} \frac{m_i\, G\, m_j}{|{\bf r}_i - {\bf r}_j|}
\equiv
 \frac{-1}{2}\sum_{i,j} m_i\, L^{-1}_{ij}\, m_j,
\end{equation}
where $m_j$ is the mass of the galaxy at site ${\bf r}_j$, and $i \neq j$.
In this ``granular'' picture, the contrast function $\delta \rho ({\bf
r}_i)/{\bar \rho}$
is equal to 1 if there is a galaxy at  ${\bf r}_i$ and is equal to $-1$ if
there is
a void. Because we have assumed all $N$ galaxies have equal mass, there is a
natural
correspondence between the $m_i$ and a two-valued $(\pm)$
``spin'' variable $s_i$:
\begin{equation}
m_i = \frac{m_o}{2}(s_i + 1).
\end{equation}
A galaxy at site ${\bf r}_i$ corresponds to ``spin-up'', while a void or
deficit corresponds
to ``spin-down''. Because of (3), we can express the Boltzmann factor and
associated
partition function of this gravitational Ising
system in terms of the spin variables as:
\begin{equation}
Z^{Grav}_{Ising}[\beta] = \sum_{\{ s_i \}} exp \left(
\frac{1}{4} \sum_{i,j} s_i\, A^{-1}_{ij}\, s_j + \frac{1}{4}\sum_j s_j h_j
\right),
\end{equation}
where $ A^{-1}_{ij}= \frac{\beta G m^2_o/2}{|{\bf r}_i - {\bf r}_j|},$ and
$h_j = \sum_i \frac{\beta m^2_0 G}{|{\bf r}_i - {\bf r}_j|}$, and we have
dropped an
irrelevent constant term. $\beta$ plays the role of an inverse temperature. The
sum in (4) is
over all possible spin configurations $s_1 = \pm 1, s_2 = \pm 1, \cdots , s_N =
\pm 1.$

To continue, we make use of the Hubbard-Stratonovich
(or gaussian) transformation \cite{Hubbard}
\begin{eqnarray}
&\pi^{N/2}& ({\rm det}A)^{-1/2}\, exp \left[\frac{1}{4} \sum_{i,j} s_i
A^{-1}_{ij} s_j \right] =\\
& & \mbox{ }
\int_{-\infty}^{\infty} \Pi^{N}_{m=1}\, dx_m \,exp \left(-\sum_{m,n} x_m A_{mn}
x_n
 + \sum_m s_m x_m \right),\nonumber
\end{eqnarray}
which maps exactly the discrete system described by the spin
variables $s_i$ onto a system
described by the continuous real
variable $-\infty < x_m < \infty$. Applying (5)  to
the partition function in (4) yields
\begin{eqnarray}
Z^{Grav}_{Ising}[\beta] &=& C\int [d\phi]\, \exp [ -\frac{\beta}{2}\int
(\phi({\bf r}) - h({\bf r}))A({\bf r},{\bf r'}) (\phi({\bf r'}) - h({\bf r'}))
 d{\bf r}d{\bf r'} \nonumber \\
&+& \int d{\bf r} \log \cosh (\beta \phi({\bf r})) ]\\ \nonumber
&\equiv& \int [d\phi] \, e^{-\beta {\cal H}[\phi,h]},
\end{eqnarray}
performing the exact configurational sum over the $\{ s_i \}$ and
after taking the continuum limit which rigorously converts the discrete
system into a continuous three dimensional 1-component scalar field theory.
The linear term in (4) results in the highly nonlinear $\log \cosh$ appearing
in (6).
$C$ is an inessential factor and
$h({\bf r})= \frac{1}{2} \int d{\bf r'}\, A^{-1}({\bf r},{\bf r'}).$
The continuous scalar field $\phi$ is the order
parameter for this system; its $n$-point Green functions correspond to the
statistically averaged product of $n$ times
the contrast evaluated at $n$-sites.
The operator $A$ in the
continuum limit is given by
\begin{equation}
A({\bf r},{\bf r'}) = \delta^{(3)}({\bf r} - {\bf r'})\,
 \frac{-\nabla^2}{2\pi \beta G m^2_0},
\end{equation}
since $\nabla^2 \frac{1}{|\bf x|} = -4\pi \delta^{(3)}({\bf x}).$

As is well known \cite{Wilson}, the connected two-point
function for the spin system and the field
theory are the same. Moreover, the unavoidable fluctuations in
the field $\phi$ result in an
{\em anomalous} dimension $\eta$ shifting the canonical dimension of the field
such that
when $|{\bf r} - {\bf r'}| \rightarrow \infty$, the two-point function for the
hamiltonian
in (6) scales as
\begin{equation}
\lim_{|{\bf r} - {\bf r'}| \rightarrow \infty} <s_i s_j> =
\lim_{|{\bf r} - {\bf r'}| \rightarrow \infty} \xi_{Gal}(|{\bf r} - {\bf r'}|)
\sim |{\bf r} - {\bf r'}|^{-(d-2+\eta)}.
\end{equation}
Here $d$ is the dimension of space $(d=3)$ and $\eta$ is the critical exponent
for the
pair correlation function whose value $(0.0198-0.064)$ (see Table 1)
differs from zero due to the \underline{fluctuations} in $\phi$.

Thus, our calculation shows that for large separations, the galaxy-galaxy
correlation
function \underline{must} scale as
\begin{equation}
\xi_{Gal}(|{\bf r}_i -{\bf r}_j|) \sim r^{-(d - 2+ \eta)},
\end{equation}
with \cite{Barber} $1.0198 \leq 1 + \eta \leq 1.064$
for $d=3$, and without taking the background expansion into account.

Thus far, our calculation has been performed for a static ensemble of galaxies,
but
the Universe is expanding and the effects of this expansion will modify the
values of the critical exponents. From the theory of dynamical critical
phenomena, it is phenomenologically known from computer simulations
that the
way time enters into the correlation
function is by a modification of its argument, viz.,
\begin{equation}
\xi(r) \rightarrow \xi(r;t) = F(r/l(t)),
\end{equation}
where $l(t) \sim t^{\zeta}$ is a function of time and here, $r$ denotes the
comoving coordinate, $r_{comoving}$, which is related to the physical
coordinate via
\begin{equation}
r_{physical} = a(t)\, r_{comoving},
\end{equation}
with $a(t)$ the scale factor for the background spacetime. From computer
simulations
in condensed matter physics, it is established that $\zeta = 1/3$ for systems
with
a conserved order parameter, while $\zeta = 1/2$ when the order parameter is
not conserved.

For a matter dominated expansion, the case we are
interested in, $a(t) = t^{2/3}$ and thus
for a given physical separation
\begin{equation}
t \sim (r_{comoving})^{-3/2}.
\end{equation}
This means that the function $l(t)$ scales as
\begin{equation}
l(t) = t^{\zeta} = (r_{comoving})^{-3\zeta/2}.
\end{equation}
Putting together these facts yields the scaling behavior of $\xi(r;t)$:
\begin{eqnarray}
\xi(r;t) &=& \left( r_{comoving}/r^{-3\zeta/2}_{comoving} \right)^{-(d - 2 +
\eta)}
   \nonumber \\
         &=& ( r_{comoving})^{-(d - 2 + \eta)(1 + 3\zeta/2)}.
\end{eqnarray}
In other words, the critical exponent resulting from the expansion of the
Universe
and dynamical critical phenomena effects is
\begin{equation}
\gamma = (d - 2 + \eta)\, \times \,(1 + 3\zeta/2).
\end{equation}
The \underline{predicted} (calculated) value for $\gamma$ is between 1.530 and
1.596
for $\zeta = 1/3$ (for a conserved order parameter) and between 1.75 to 1.862
for
$\zeta = 1/2$ (non-conserved order parameter). Table 1 displays these
exponents for both the static and expanding cases (C, NC, denote conserved and
nonconserved order parameter, respectively). The static values are taken from
Ref[9]. The last four entries refer to expansions in powers of
$\epsilon = 4 - d$.
These values are to be compared
with the values inferred from current galaxy catalogs, which range from 1.5 for
the APM survey, to 1.8 for the Lick survey \cite{APM,Lick}.

\begin{table}[t]
\caption{Calculated values for the galaxy-galaxy correlation function critical
exponent}
\vspace{0.4cm}
\begin{center}
\begin{tabular}{|c|c|c|c|}
\hline
Method of Calculation& $\gamma_{Static}$ & $\gamma^{C}_{Expanding}$
 & $\gamma^{NC}_{Expanding}$ \\
\hline
Series estimates& $1.056 \pm 0.008$ & $1.584 \pm 0.012$ & $1.848 \pm 0.014$ \\
$O(\epsilon)$   & 0  & 1.5  & 1.75  \\
$O(\epsilon^2)$ & 1.0198 & 1.530 & 1.785 \\
$O(\epsilon^3)$ & 1.037  & 1.555 & 1.815 \\
$O(\epsilon^4)$ & 1.029  & 1.543 & 1.801 \\
\hline
\end{tabular}
\end{center}
\end{table}

We have thus accomplished what we set out to do, namely, provide
a first-principles
calculation of the galaxy-galaxy correlation function
using standard methods of
statistical mechanics and the theory of critical phenomena.
Naturally, one might (and should) ask {\it why} this
calculation is successful, in other words,
what does a system of galaxies have to do with a spin model?
 The answer is provided
by the universality hypothesis, according to which rather
distinct physical systems
will exhibit {\em identical} critical behavior provided the systems in question
possess the same space dimensionality and have order parameters with the
same dimensionality (or number of independent components). We see that this is
the case
here: $d=3$ for both systems, while the scalar field and the magnetization are
one-component order parameters for the galaxy system and the 3d Ising model,
respectively. In other words, both systems belong to the same universality
class.
Thus, we must expect them to have the same values for
their critical exponents.

\section*{Acknowledgements}
D.H. would like to thank the organizers
for the invitation to participate in this
Workshop and for their kind and generous hospitality.
Thanks are also due to the participants
for their enthusiastic response to this talk, as evidenced by their many
thoughtful
comments and questions.


\begin{thebibliography}{99}
\bibitem{footnote1} As we shall demonstrate below, the galaxy-galaxy
correlation
function can be expressed as a scalar field propagator. In units of length, a
scalar
field in $d$-dimensional space has a canonical dimension of $-(d/2 - 1)$. The
above
statement about $\gamma$ follows at once.
\end{thebibliography}
\end{document}